\documentclass[11pt]{article} 

\usepackage[utf8]{inputenc} 
\usepackage[english]{babel}
\usepackage{hyperref}
\usepackage{latexsym}
\usepackage{amsfonts}
\usepackage{times}
\usepackage{xcolor}
\usepackage{graphicx}
\usepackage{xspace}
\usepackage{algorithmicx}
\usepackage{natbib}

\newcommand*{\affaddr}[1]{#1} 
\newcommand*{\affmark}[1][*]{\textsuperscript{#1}}
\newcommand*{\email}[1]{\texttt{#1}}

\title{\LARGE\textbf{Temporal Bibliographic Networks}}

\author{
Vladimir Batagelj\affmark[1,2,3]  Daria Maltseva\affmark[1] \\
ORCID: 0000-0002-0240-9446, 0000-0003-1789-1711\\
\affaddr{\affmark[1] National Research University Higher School of Economics,\\ Myasnitskaya, 20, 101000 Moscow, Russia.}\\
\affaddr{\affmark[2]Institute of Mathematics, Physics and Mechanics,\\ Jadranska 19, 1000 Ljubljana, Slovenia}\\
\affaddr{\affmark[3]University of Primorska, Andrej Marušič Institute, 6000 Koper, Slovenia}\\ 
\email{vladimir.batagelj@fmf.uni-lj.si}\\
\email{d\_malceva@mail.ru}
}

\newcommand{\keyw}[1]{\textcolor{red}{\emph{#1}}}

\newcommand{\WA}{\mathbf{W\!\!A}}

\newcommand{\WK}{\mathbf{W\!K}}

\newcommand{\WJ}{\mathbf{W\!J}}
\newcommand{\Ci}{\mathbf{Cite}}

\newcommand{\RR}{\Bbb{R}}
\newcommand{\NN}{\Bbb{N}}

\newcommand{\network}[1]{\mathcal{#1}}
\newcommand{\vertices}[1]{\mathcal{#1}}
\newcommand{\edges}[1]{\mathcal{#1}}

\newcommand{\functions}[1]{\mathcal{#1}}

\newcommand{\outdeg}{\mbox{outdeg}}

\newcommand{\func}[1]{\textit{#1}}

\newcommand{\Time}{\mathcal{T}}
\newcommand{\cmdkey}{\raisebox{-.025em}{\includegraphics[height=.7em]{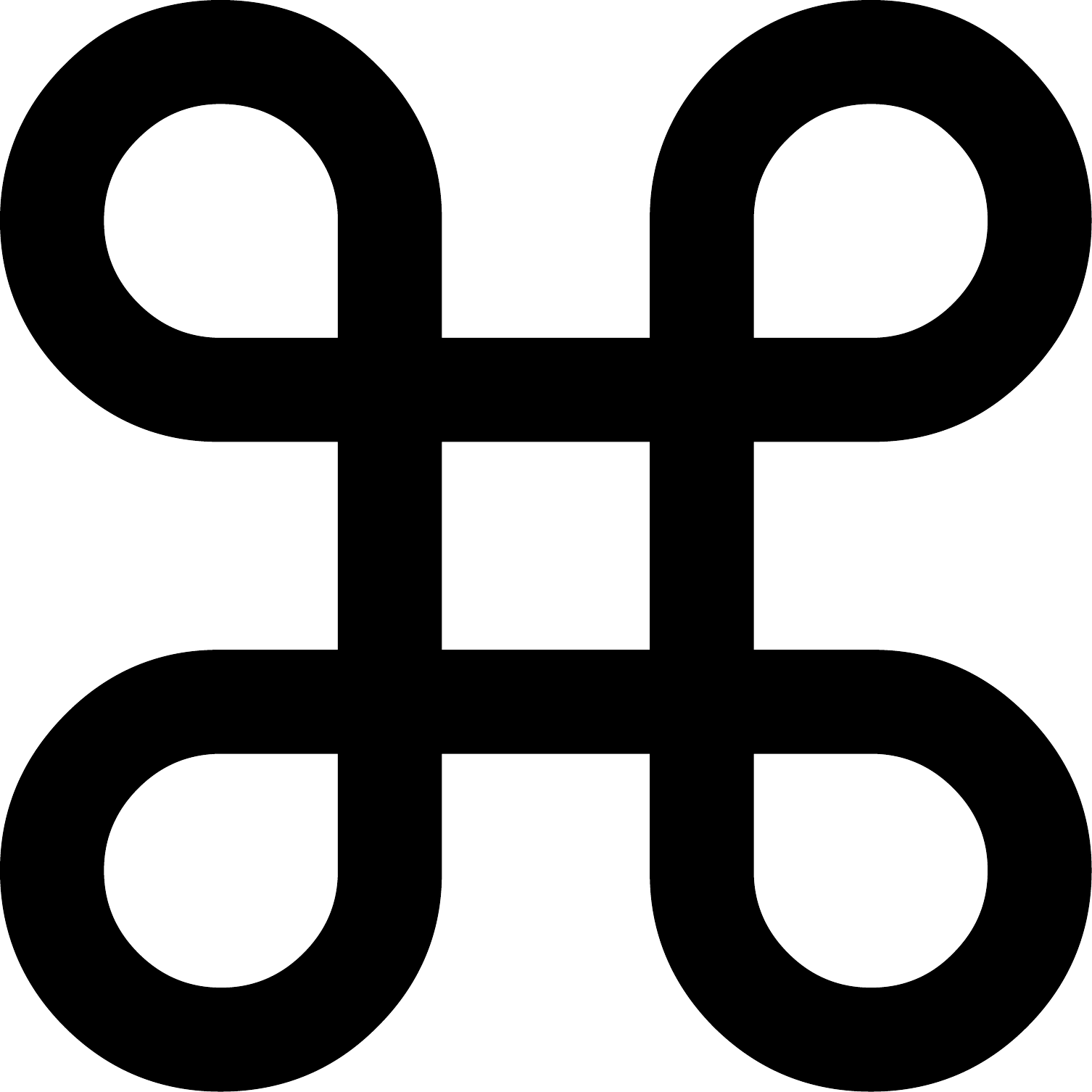}}}
\newcommand{\Mw}{\mathop{\raisebox{-1.5pt}{\mbox{$\Box$\kern-.55em\raisebox{2.5pt}{{\tiny $r$}}\kern2.9pt}}}}
\newcommand{\Mv}{\mathop{\raisebox{-1.5pt}{\mbox{$\Box$\kern-.55em\raisebox{2.5pt}{{\tiny $h$}}\kern2.9pt}}}}

\newcommand{\Remark}[1]{\ifodd\value{page} \normalmarginpar
 \else \reversemarginpar \fi \marginpar{{\footnotesize #1}} }

\newcommand{\clock}{\count254=\time \divide\count254 by 60
 \count255=\count254 \multiply\count255 by -60
 \advance\count255 by \time
 \ifnum\count254<10 0\fi\number\count254\,:\,%
 \ifnum\count255<10 0\fi\number\count255}

\graphicspath{{./}{./pics/}}

\begin{document}

\hypersetup{pdfauthor={V. Batagelj, D. Maltseva}}
\hypersetup{pdftitle={Temporal Bibliographic Networks}}

\maketitle

\begin{abstract}
We present two ways (instantaneous and cumulative) to transform bibliographic networks, using the works' publication year, into corresponding temporal networks based on temporal quantities. We also show how to use the addition of temporal quantities to define interesting temporal properties of nodes, links and their groups thus providing an insight into evolution of bibliographic networks. Using the multiplication of temporal networks we obtain different derived temporal networks providing us with new views on studied networks. The proposed approch is illustrated with examples from the collection of bibliographic networks on peer review. 
\\[4pt]
\\[4pt]
\\[4pt]

\end{abstract}


\section{Introduction}

From data collected from bibliographic databases (WoS, Scopus, Google scholar, Bibtex, etc.) we can construct different bibliographic networks. For example using the program WoS2Pajek we obtain from data collected from WoS 
the following two-mode networks: 
the authorship network $\WA$ on works $\times$ authors,
the journalship network $\WJ$ on  works $\times$ journals,
the keywordship network $\WK$ on works  $\times$ keywords, and
the (one-mode) citation network $\Ci$ on works.
We obtain also the following node properties: 
the partition $year$ of works by publication year, 
the $DC$ partition distinguishing between works with complete description ($DC[w]=1$) and the cited only works ($DC[w]=0$), and the vector of number of pages $N\!P$.
Analyzing these networks we can get distributions of frequencies of different units (authors, journals, keywords) describing overall properties of networks. We can also identify the most important units \citep{ZB}. An important tool in the analysis of linked (collections of) networks is the network multiplication that produces derived networks linking not directly linked sets of units -- for example, the network $ \mathbf{AK} = \WA^T \cdot \WK$ links authors to keywords \citep{OnBibl}.

A more detailed insight in the evoultion of bibliographic networks is enabled by considering also the temporal information. In the paper \cite{tq} a longitudinal approach to analysis of temporal networks based on temporal quantities was presented. It is an alternative to the traditional cross-sectional approach \citep{TN}. In this paper we show how to apply the proposed approach to temporal bibliographic networks. It can be used also in other similar contexts.

First we describe two ways how the year of publication can be combined with traditional bibliographic networks to get their temporal versions -- the instantaneous and the cumulative. Afterward we present different ways to analyze these networks and networks derived from them using network multiplication. 

The proposed approch is illustrated with examples on networks from the collection of bibliographic networks on peer review \citep{PeerRew} on works with complete descriptions. The sizes of different sets of units are as follows:
$| W | = 22104$,  $| A | = 62106$,  $| J | = 6716$, and  $| K | = 36275$.

\section{Temporal networks}

A \keyw{temporal network}
$\network{N}_T =(\vertices{V},\edges{L}, \Time,\functions{P},\functions{W})$
is obtained by attaching the \keyw{time}, $\Time$, to an ordinary network where
$\Time$ is a set of  \keyw{time points}, $t \in \Time$.\medskip

In a temporal network, nodes $v \in \vertices{V}$ and links $l \in \edges{L}$
are not necessarily present or active in all time points.
Let $T(v)$, $T \in \functions{P}$, be the \keyw{activity set} of time points for node $v$ and  $T(l)$, 
$T \in \functions{W}$,   the activity set of time points for link $l$.\medskip

Besides the presence/absence of nodes and links also their properties can change
through time.


\subsection{Temporal quantities}

We introduce a notion of a \keyw{temporal quantity}
\[  a(t) = \left\{\begin{array}{ll} 
                a'(t) & t \in T_a \\
                \cmdkey & t \in \Time \setminus T_a
             \end{array}\right. \]
where $T_a$ is the \keyw{activity time set} of $a$, $a'(t)$ is the value of $a$ in an instant $t \in T_a$, and  \cmdkey{} denotes the value \keyw{undefined}.

We assume that the values of temporal quantities belong to a set $A$ which is
a \keyw{semiring} $(A,+,\cdot,0,1)$ for binary operations $+ : A\times A \to A$ and
$\cdot : A\times A \to A$. The semiring $(\RR_0^+,+,\cdot,0,1)$ where $+$ is addition and $\cdot$ is multiplication of numbers is called a \keyw{combinatorial} semiring. For solving the shortest path problems on networks the semiring 
$(\RR_0^+\cup\{\infty\},\min,+,\infty,0)$ is used \citep{SemiEx}.

We can extend both operations to the set
$A_{\scriptsize\cmdkey} = A \cup \{\cmdkey\}$ by requiring that for all $a \in A_{\scriptsize\cmdkey}$
it holds
\[ a + \cmdkey = \cmdkey + a = a \quad \mbox{and} \quad
   a \cdot \cmdkey = \cmdkey \cdot a = \cmdkey . \]
The structure $(A_{\scriptsize\cmdkey},+,\cdot,\cmdkey,1)$ is also a semiring.

Let $A_{\scriptsize\cmdkey}(\Time)$ denote the set of all temporal quantities
over $A_{\scriptsize\cmdkey}$ in time $\Time$. To extend the operations to
networks and their matrices we first define the \keyw{sum} (parallel links)
$ a + b $  as
\[ (a+b)(t) =  a(t) + b(t) \quad \mbox{ and } \quad T_{a + b} = T_a \cup T_b .\]
The \keyw{product} (sequential links) 
$ a \cdot b $ is defined as             
\[ (a \cdot b)(t) =  a(t) \cdot b(t) \quad \mbox{ and } \quad T_{a \cdot b} = T_a \cap T_b . \]

Let us define temporal quantities $\mathbf{0}$ and $\mathbf{1}$ with
requirements $\mathbf{0}(t) = \cmdkey$ and $\mathbf{1}(t) = 1$ for all
$t \in \Time$. Again, the structure 
$(A_{\scriptsize\cmdkey}(\Time),+,\cdot,\mathbf{0},\mathbf{1})$ is a
semiring.

To produce a software support for computation with temporal quantities we limit it to temporal quantities that can be described as a sequence of disjoint time intervals with a constant value
\[ a = [(s_i,f_i,v_i)]_{i \in 1..k} \] 
where $s_i$ is the starting time and $f_i$ the finishing time of the $i$-th time interval $[s_i, f_i)$, $s_i < f_i$ and $f_i \leq s_{i+1}$, and $v_i$ is the value of $a$ on this interval. Outside the intervals the value of temporal quantity $a$ is undedined, $\cmdkey$. Therefore 
\[ T_a = \bigcup_{i \in 1..k} [s_i, f_i)  . \]

\begin{figure}
 \begin{center}
  \small
    \begin{tabular}{rl}
  \raisebox{10mm}{$a+b$ :} &
   \includegraphics[width=60mm,viewport=140 80 580 260,clip=]{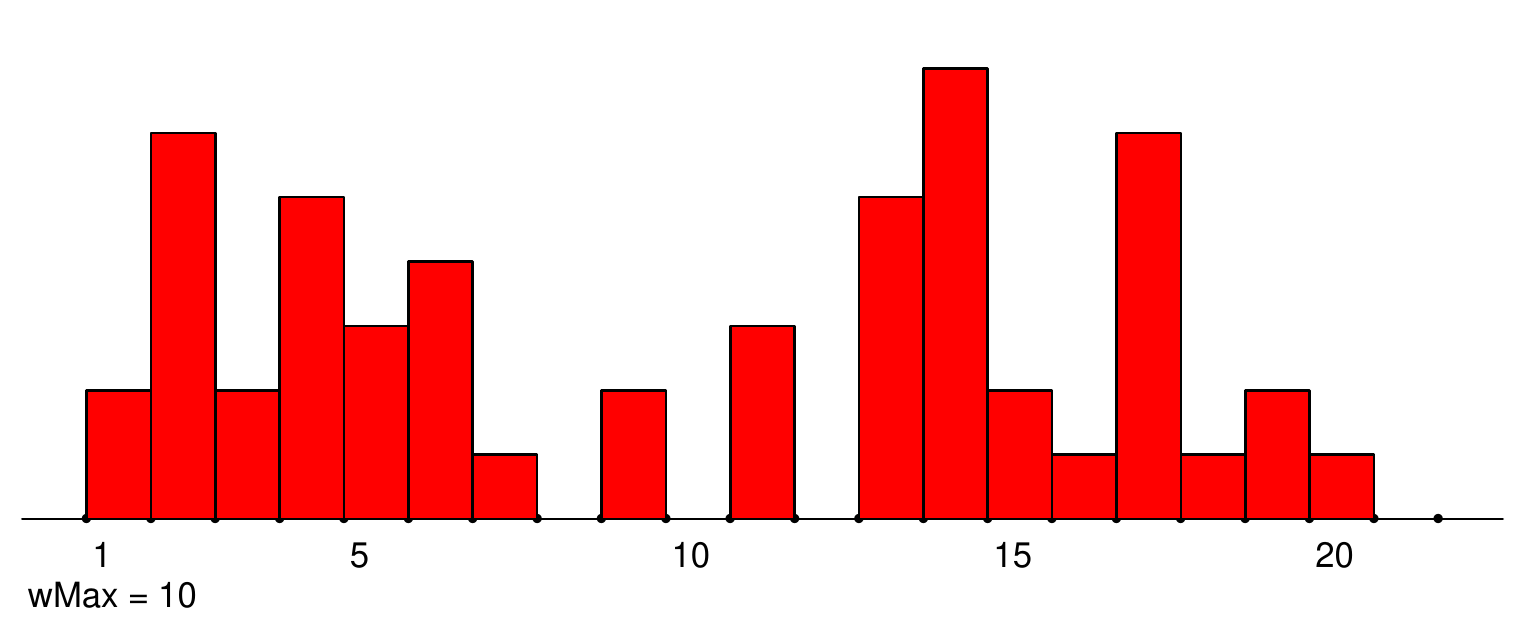}\\
  \raisebox{10mm}{$a$ :} &
   \includegraphics[width=60mm,viewport=140 80 580 220,clip=]{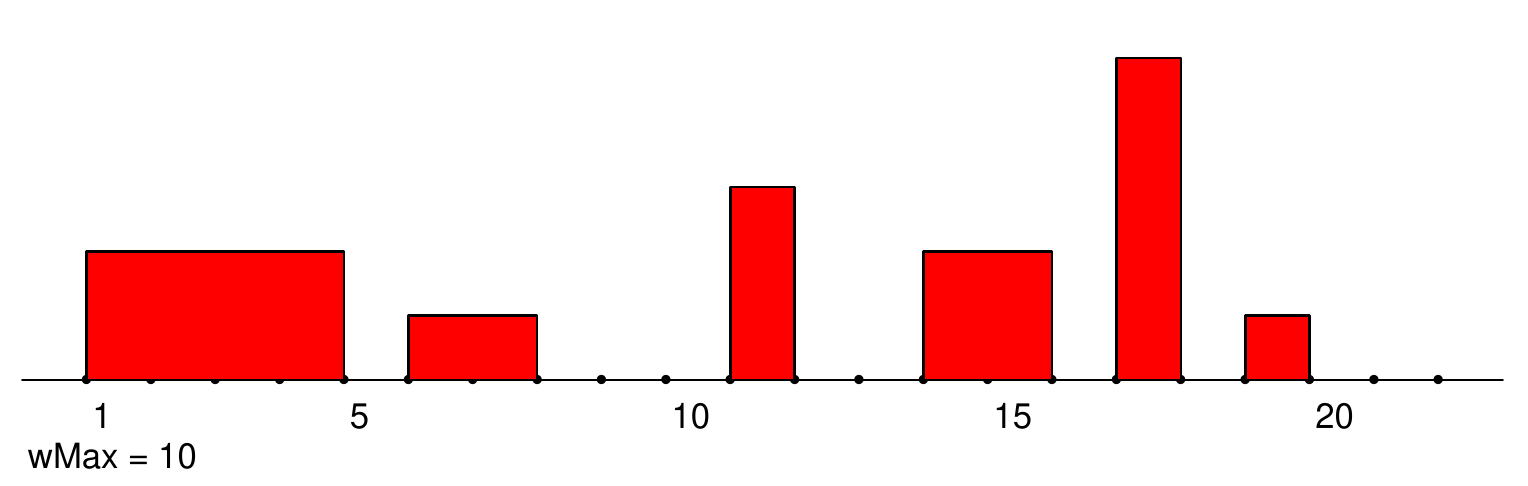}\\
  \raisebox{10mm}{$b$ :} & 
  \includegraphics[width=60mm,viewport=140 80 580 220,clip=]{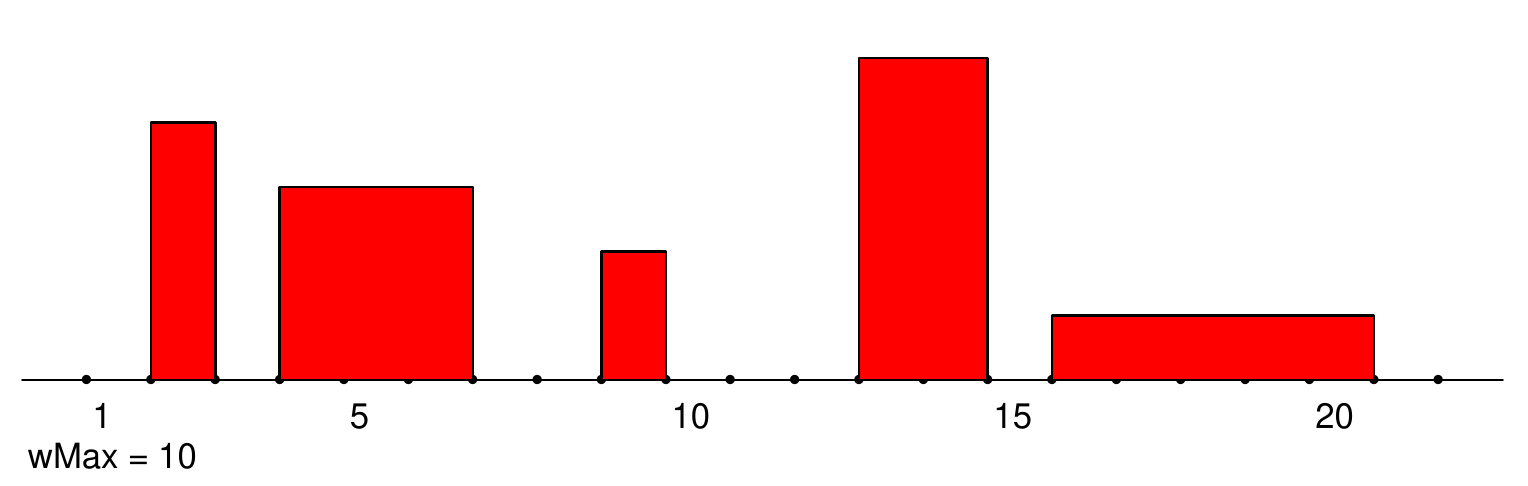}\\
  \raisebox{10mm}{$a\cdot b$ :} &
   \includegraphics[width=60mm,viewport=140 80 580 305,clip=]{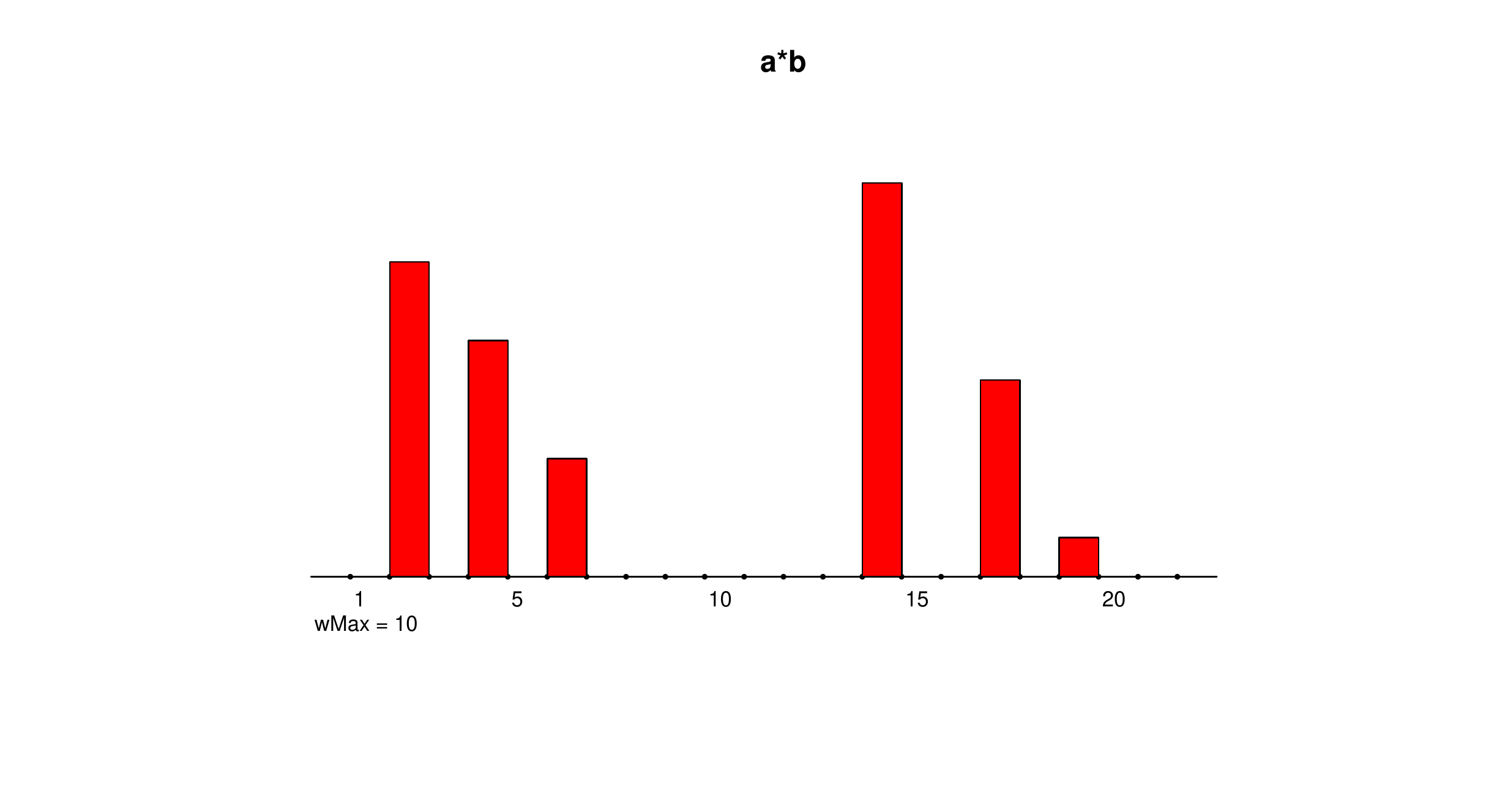}
    \end{tabular}
  \caption{Sum and product of temporal quantities\label{sp}}  
 \end{center}
\end{figure}

To illustrate both operations let us consider temporal quantities $a$ and $b$ \citep{tq}:
\begin{eqnarray*}
a  &=& [(1, 5, 2), (6, 8, 1), (11, 12, 3), (14, 16, 2),  (17, 18, 5), (19, 20, 1)]\\
b  &=&  [(2, 3, 4), (4, 7, 3), (9, 10, 2), (13, 15, 5), (16, 21, 1)]
\end{eqnarray*}
The following are the sum $s = a+b$ and the product $p = a\cdot b$ of temporal quantities $a$ and $b$ over combinatorial semiring.
\begin{eqnarray*}
s  &=& [(1, 2, 2), (2, 3, 6), (3, 4, 2), (4, 5, 5), (5, 6, 3), 
     (6, 7, 4), (7, 8, 1), (9, 10, 2), (11, 12, 3), \\
 & &    (13, 14, 5), (14, 15, 7), (15, 16, 2), (16, 17, 1), 
     (17, 18, 6), (18, 19, 1), (19, 20, 2), (20, 21, 1)]\\
p  &=& [(2, 3, 8), (4, 5, 6), (6, 7, 3), (14, 15, 10), (17, 18, 5), (19, 20, 1)] 
\end{eqnarray*}
They are visually displayed in  Figure~\ref{sp}.

To support  computations with temporal quantities and analysis of temporal networks based on them the Python libraries TQ and Nets were developed \citep{ref4}. They were used in analyses presented in this paper. In the examples we used a collection of bibliographic networks on peer review from \citet{PeerRew}.


\subsection{Temporal affiliation networks \label{teco}}

Let the binary \keyw{affiliation} matrix $\mathbf{A}=[a_{ep}]$ describe a two-mode network on the set of
events $E$ and the set of of participants $P$:
\[  a_{ep} = \left\{\begin{array}{ll} 
                1 & p \mbox{ participated at  the event } e \\
                0 & \mbox{otherwise}
             \end{array}\right. \]
The function $d: E \to \Time$ assigns to each event $e$ the date $d(e)$
when it happened. Assume $\Time = [\func{first}, \func{last}]\subset \NN$. Using these data we can construct two
temporal affiliation matrices:
\begin{itemize}
\item \textbf{instantaneous} $\mathbf{Ai}=[ai_{ep}]$, where
\[  ai_{ep} = \left\{\begin{array}{ll} 
                [(d(e),d(e)+1,1)] & a_{ep} = 1 \\
                \lbrack\ \rbrack & \mbox{otherwise}
             \end{array}\right. \]
\item \textbf{cumulative} $\mathbf{Ac}=[ac_{ep}]$, where            
\[  ac_{ep} = \left\{\begin{array}{ll} 
                [(d(e),last+1,1)] & a_{ep} = 1 \\
                \lbrack\ \rbrack  & \mbox{otherwise}
             \end{array}\right. \]
\end{itemize}

In general a temporal quantity $a$ is called \keyw{cumulative} iff it has for $t, t' \in \Time$ the property
\[  t \in T_a  \ \land \ t' > t \ \   \Rightarrow \  \  t' \in T_a   \ \land \   a(t') \geq a(t)  \]
A sum and product (over combinatorial semiring)  of  cumulative temporal quantities are cumulative temporal quantities.

For a temporal quantity $a = [(s_i,f_i,v_i)]_{i \in 1..k}$ its  \keyw{cumulative} $cum(a)$ is defined as
\[ cum(a) =  [(s_i,s_{i+1},V_i)]_{i \in 1..k} \]
where $s_{k+1} = last$ and $V_i = \sum_{j=1}^i v_j$ .

A temporal network is cumulative for a weight $w$ iff all its values are cumulative.

The Python code for creating temporal networks from Pajek files for the peer review data is given in Appendix~A.1.


\subsection{Temporal properties  \label{teco}}

Let $\mathbf{N}$ be a temporal network on $E \times P$. On it we can define some interesting temporal quantiries such as
\keyw{in-sum}:
\[ iS(\mathbf{N},p) = \sum_{e \in E} n_{ep} \]
and \keyw{out-sum}:
\[ oS(\mathbf{N},e) = \sum_{p \in P}  n_{ep} \]

In a special case where  $\mathbf{N} \equiv \WA\mathbf{i}$ we get the  \keyw{productivity of an author} $a$
\[ pr(a) =  iS(\WA\mathbf{i},a) =  \mbox{ number of publications of the author $a$ by year}\]
and for $\mathbf{N} \equiv \WA\mathbf{c}$ we get the  \keyw{cumulative productivity of an author} $a$
\[ cpr(a) =   iS(\WA\mathbf{c},a) = \mbox{ cumulative number of publications of the author $a$ by year.} \] 
It holds $cpr(a) = cum(pr(a))$.

The productivity of an author can be extended  to the \keyw{productivity of a group of authors} $C$
 \[  pr(C) = \sum_{a \in C} pr(a) = \sum_{a \in C}  iS(\WA\mathbf{i},a) \]
 There is a problem with the productivity of a group. In the case when two authors from a group co-authored the same paper it is counted twice. To account for a ``real'' contribution of each author the fractional approach is used. It is based on normalized networks (matrices) -- in the case of co-authorship on $n(\WA) = \WA\mathbf{n} = [ wan_{wa}]$
 \[ wan_{wa} = \frac{wa_{wa}}{\max(1,\outdeg_\WA(w))} . \]
This leads to the \keyw{fractional productivity of an author} a
\[ fpr(a) =  iS(\WA\mathbf{ni},a) =  \mbox{ fractional contribution of publications of the author $a$ by year}\]

\subsubsection{Example: Temporal properties in networks on  peer review  \label{exteco}} 

\begin{figure}
\begin{center}
\includegraphics[width=100mm]{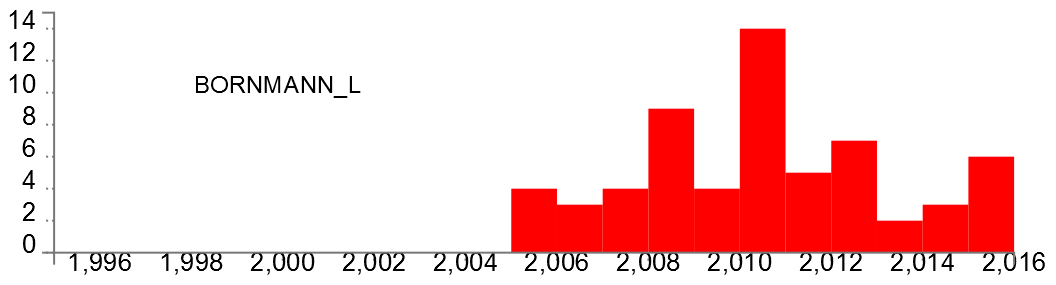}\\[2mm]
\includegraphics[width=100mm]{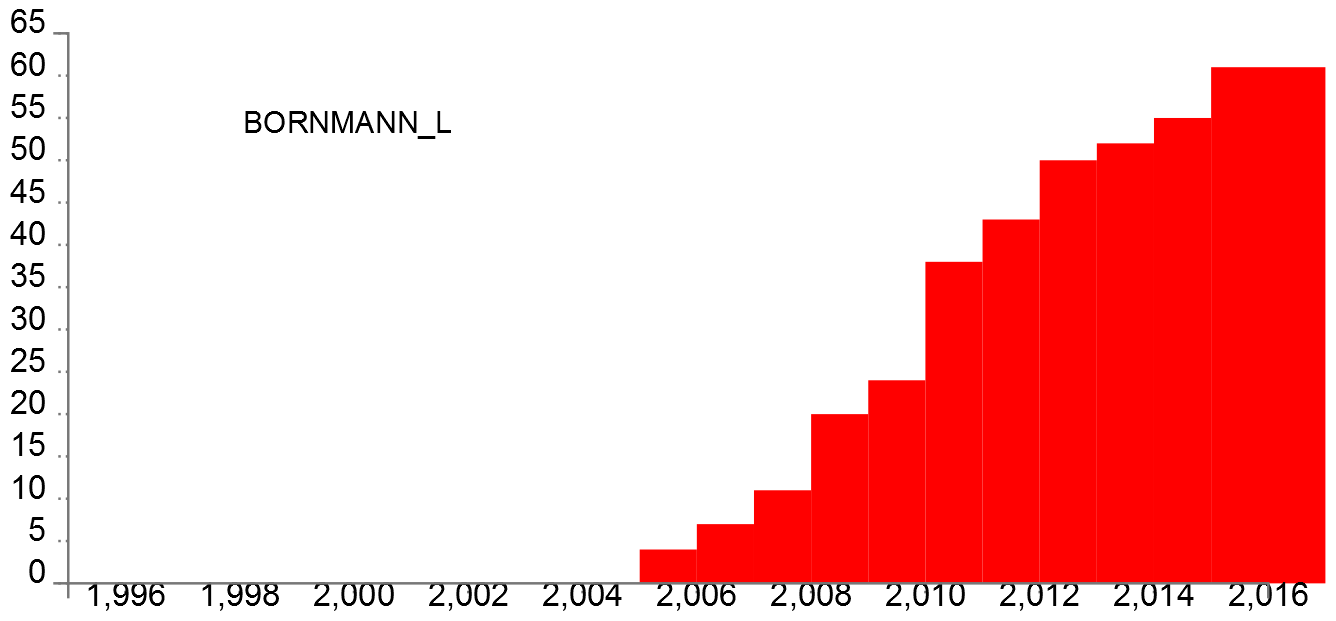}\\[2mm]
\includegraphics[width=100mm]{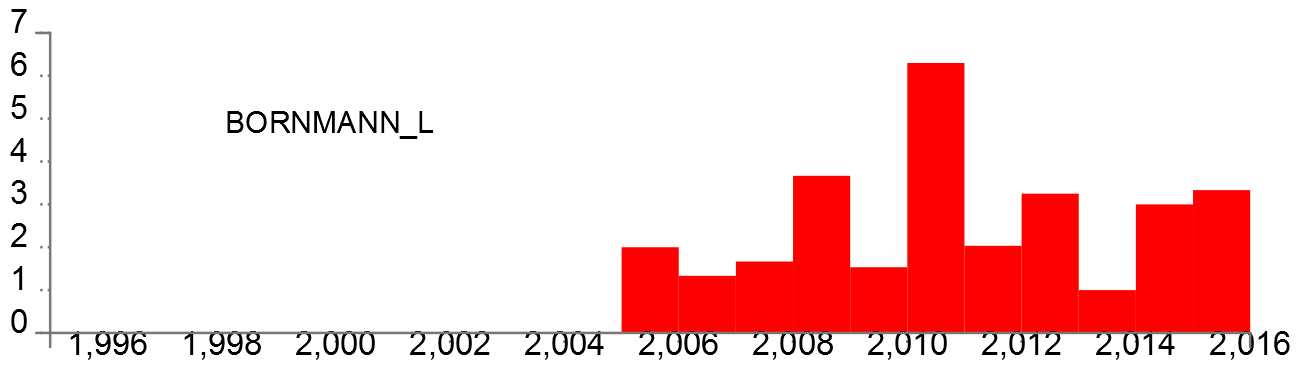}
\caption{Productivity, cumulative productivity and fractional productivity of Lutz Bornmann\label{LB}}
\end{center}
\end{figure}

In the analysis of the ordinary authorship network $\WA$ we get that  Lutz Bornmann is the author who wrote the largest number, 61,  of works on peer review \citep{PeerRew}.To see the dynamics of his publishing we compute his productivity
\begin{eqnarray*}
pr  &=& [(2005, 2006, 4), (2006, 2007, 3), (2007, 2008, 4), (2008, 2009, 9),\\
      & &  (2009, 2010, 4), (2010, 2011, 14), (2011, 2012, 5), (2012, 2013, 7),\\
      & &  (2013, 2014, 2), (2014, 2015, 3), (2015, 2016, 6)]
\end{eqnarray*}
see the top of Figure~\ref{LB}.  The corresponding cumulative productivity is
\begin{eqnarray*}
cpr &=& [(2005, 2006, 4), (2006, 2007, 7), (2007, 2008, 11), (2008, 2009, 20),\\
      & &   (2009, 2010, 24), (2010, 2011, 38), (2011, 2012, 43), (2012, 2013, 50), \\
      & &   (2013, 2014, 52), (2014, 2015, 55), (2015, 2017, 61)]
\end{eqnarray*}
see the mid of Figure~\ref{LB}. Note that $cpr = cum(pr)$: $7=4+3$, $11=4+3+4$, \ldots 

The fractional productivity of   Lutz Bornmann is
\begin{eqnarray*}
fpr &=& [(2005, 2006, 2.0), (2006, 2007, 1.333), (2007, 2008, 1.667), (2008, 2009, 3.667),\\
     & &   (2009, 2010, 1.533), (2010, 2011, 6.3), (2011, 2012, 2.033), (2012, 2013, 3.25), \\
     & &   (2013, 2014, 1.0), (2014, 2015, 3.0), (2015, 2016, 3.333)] 
\end{eqnarray*}
see the bottom of Figure~\ref{LB}. For the Python code see Appendix~A.2.

\begin{figure}
\begin{center}
\includegraphics[width=100mm]{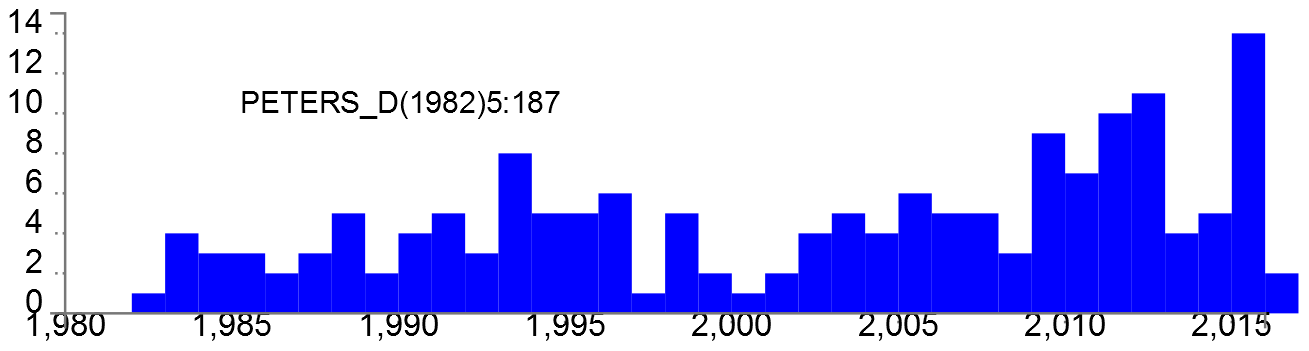}\\[2mm]
\includegraphics[width=100mm]{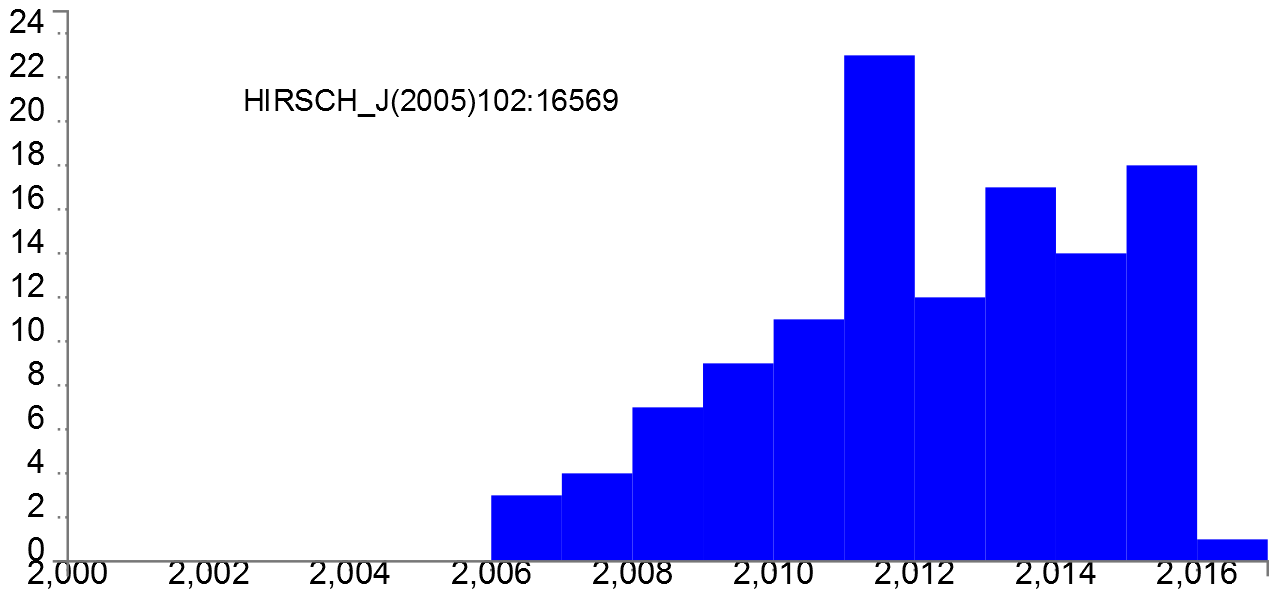}
\caption{Citations to Peters and Hirsch\label{PH}}
\end{center}
\end{figure}

In the citation network $\mathbf{Cite}$ for the peer review bibliography the most cited, 164, paper is 
Peters, D. P., Ceci, S. J. (1982). Peer-review practices of psychological journals: The fate of published articles, submitted again. Behavioral and Brain Sciences, 5(2), 187-255. The temporal quantity $citP =  iS(\Ci\mathbf{i},\mbox{PETERS\_D(1982)5:187})$ describes the number of citations to this paper through years.
\begin{eqnarray*}
citP  &=& [(1982, 1983, 1), (1983, 1984, 4), (1984, 1986, 3), (1986, 1987, 2), \\
      & &   (1987, 1988, 3), (1988, 1989, 5), (1989, 1990, 2), (1990, 1991, 4),\\
      & &   (1991, 1992, 5), (1992, 1993, 3), (1993, 1994, 8), (1994, 1996, 5), \\
      & &   (1996, 1997, 6), (1997, 1998, 1), (1998, 1999, 5), (1999, 2000, 2), \\
      & &   (2000, 2001, 1), (2001, 2002, 2), (2002, 2003, 4), (2003, 2004, 5), \\
      & &  (2004, 2005, 4), (2005, 2006, 6), (2006, 2008, 5), (2008, 2009, 3), \\
      & &  (2009, 2010, 9), (2010, 2011, 7), (2011, 2012, 10), (2012, 2013, 11),\\
      & &  (2013, 2014, 4), (2014, 2015, 5), (2015, 2016, 14), (2016, 2017, 2)]
\end{eqnarray*}
See the top of Figure~\ref{PH}.

Another well known paper is Hirsch, J.E. (2005). An index to quantify an individual's scientific research output. Proc Natl Acad Sci U S A. 2005 Nov 15;102(46):16569-72 with 119 citations and  $citH =  iS(\Ci\mathbf{i},\mbox{HIRSCH\_J(2005)102:16569})$
\begin{eqnarray*}
citH  &=& [(2006, 2007, 3), (2007, 2008, 4), (2008, 2009, 7), (2009, 2010, 9), \\
      & &    (2010, 2011, 11), (2011, 2012, 23), (2012, 2013, 12), (2013, 2014, 17),\\
      & &    (2014, 2015, 14), (2015, 2016, 18), (2016, 2017, 1)]
\end{eqnarray*}
See the bottom of Figure~\ref{PH}.  For the Python code see Appendix~A.3.

Similarly we could look at the number of works by year $wy = \sum_{j \in J} iS(\WJ\mathbf{i},j)$,  the popularity of a keyword $k$:  $pop(k) = iS(\WK\mathbf{i},k)$, etc.

\subsubsection{Example: main journals publishing on peer review \label{main}}

\begin{figure}
\begin{center}
\includegraphics[width=111mm,viewport=5 14 388 390,clip=]{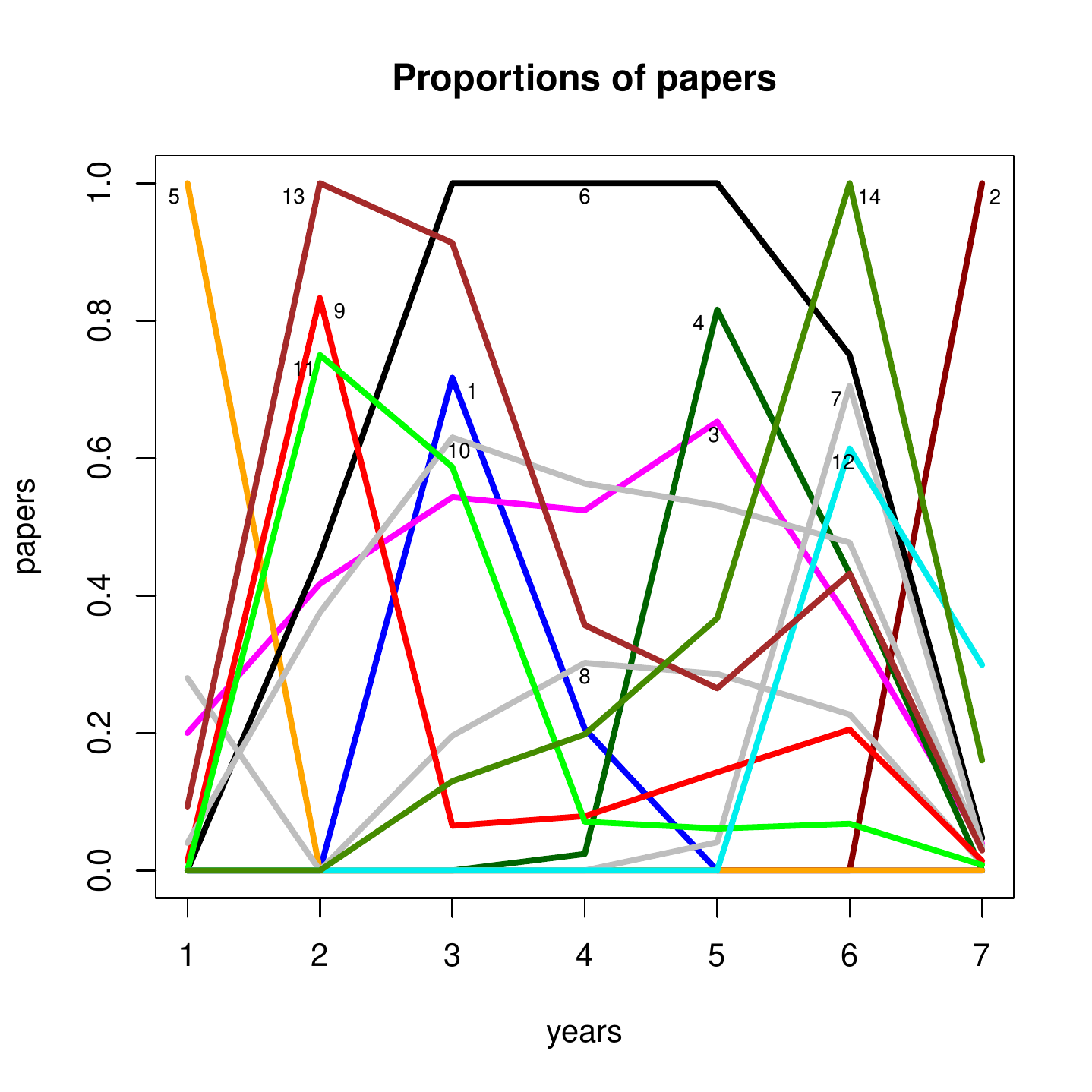} 
\footnotesize
\begin{tabular}[b]{rl}
  i &  journal   \\ \hline
 1 & BEHAV BRAIN SCI      \\
 2 & BMJ OPEN              \\
 3 & BRIT MED J            \\
 4 & CUTIS                   \\
 5 & J ASSOC OFF AGR CHEM  \\
 6 & JAMA-J AM MED ASSOC   \\
 7 & J SEX MED             \\
 8 & LANCET  \\
 9 & MED J AUSTRALIA \\
10 & NATURE \\
11 & NEW ENGL J MED \\
12 & PLOS ONE  \\
13 & SCIENCE \\
14 & SCIENTOMETRICS \\
\hline
\vspace*{25mm}
\end{tabular}
\normalsize
\caption{The most important  journals\label{Jours}}
\end{center}
\end{figure}

To identify the main journals publishing on peer review, see Appendix~\ref{important}, we determined first the temporal inSums $Jt$ in the network $\mathbf{JWi}$  for all journals. An entry $Jt(j) = iS(\mathbf{JWi},j)$ contains the temporal quantity counting the number of papers on peer review published in  the journal $j$ in each year. Because most of frequencies are small (one digit numbers) we decided to change the time scale (granularity) to time intervals:  1: 1900-1970, 2: 1971-1980, 3: 1981-1990, 4: 1991-2000, 5: 2001-2005, 6: 2006-2010, 7:2011-2015. The recoded table $Jt$ is labeled $Jr$. For the table $Jr$ we determined for each time interval three the most frequently used journals -- they are listed on the right side of Figure~\ref{Jours}. The corresponding data were exported as \texttt{journals.csv} and visualized using R. The picture on the left side presents the trajectories of relative importance (journal's frequency divided with the maximum frequency on the interval) for the selected journals.

The papers on peer review (refereeing) published till 1970 appeared most often in J ASSOC OFF AGR CHEM. Till 2005 the dominant journals were JAMA, SCIENCE, NATURE, BRIT MED J, and LANCET (general medical and science journals). In the period 2006-2010 the leading role was overtaken by a specialized journal SCIENTOMETRICS. In the last period 2011-2015 the primate is shifted to the mega-journals BMJ OPEN and PLOS ONE \citep{mega}. Note that the frequencies for SCIENTOMETRICS are 3: 6, 4: 25, 5:18, 6: 44, 7: 78 and in the period 2011-2015 489 papers on peer review were published in BMJ OPEN.

\section{Network multiplication and derived networks}



Let $\mathbf{A}$ on $A \times P$ and $\mathbf{B}$ on $P \times B$ be (matrices of linked two-mode)  networks.
Their \keyw{product} network is determined by a matrix $\mathbf{C} = [c_{i,j}]$ on $A \times B$ of the product of corresponding matrices
\[ \mathbf{C} = \mathbf{A} \cdot \mathbf{B} \]
where 
\[ c_{i,j} = \sum_{p \in P} a_{i,p} \cdot b_{p,j} \]
For details see \citet{OnBibl}. 

Network multiplication is very important in network analysis of collections of linked networks because it enables us to construct different \keyw{derived} networks. For example, in analysis of bibliographic networks the network
\[ \mathbf{AK} = \mathbf{WA} ^ T \cdot \mathbf{WK} \] 
links authors to keywords: the  weight of the arc from the node $a$ to the node $k$  is equal to the number of works in which the author $a$ used the  keyword $k$. \\
The coauthorship network $ \mathbf{Co}$ is obtained as
\[ \mathbf{Co} = \mathbf{WA}^T \cdot \mathbf{WA} \] 
The weight $co_{ab}$  is equal to total number of works authors $a$ and $b$ wrote together. \\
The network of normalized citations between authors 
\[ \mathbf{CiteAn} = n(\WA) ^ T \cdot n(\mathbf{Cite}) \cdot n(\WA) \]  
The weight $citean_{uv}$ is equal to the  fractional contribution  of citations from works coauthored by $u$ to works coauthored by $v$.
Etc.

The network (matrix) multiplication can be straightforwardly extended to temporal networks.


\subsection{Multiplication of temporal networks  \label{mulins}}

Let $\mathbf{A}$ on $A \times P$ and $\mathbf{B}$ on $P \times B$ be (matrices of) co-occurence networks. Then  $\mathbf{C} = \mathbf{A} \cdot \mathbf{B}$ is a temporal network on  $A \times B$. What is its meaning? Consider the value of its item in an instant $t$
\[ c_{ij}(t) = \sum_{p\in P} a_{ip}(t)^T \cdot b_{pj}(t) = \sum_{p\in P} a_{pi}(t) \cdot b_{pj}(t)\]
For $c_{ij}(t)$ to be defined (different from $\cmdkey$) there should be at least one $p\in P$ such that $a_{pi}(t)$ and  $b_{pj}(t)$ are both defined, i.e. $t \in T_{a_{pi}} \cap  T_{b_{pj}}$. Then there exists $g_{pi}$ such that $(s_{g_{pi}},f_{g_{pi}},v_{g_{pi}}) \in a_{pi}$, $t \in [ s_{g_{pi}},f_{g_{pi}} )$, and $a_{pi}(t) = v_{g_{pi}}$. Similarly $b_{pj}(t) = v_{h_{pj}}$. Therefore
\[ c_{ij}(t) = \sum_{p: t \in T_{a_{pi}} \cap  T_{b_{pj}}} v_{g_{pi}} \cdot v_{h_{pj}} \]

For binary instantaneous two-mode networks $\mathbf{A}$ and $\mathbf{B}$ the value
$c_{ij}(t)$ of the product $\mathbf{C} = \mathbf{A} \cdot \mathbf{B}$ is equal to the number
of different members of $P$ with which both $i$ and $j$ have contact in the instant $t$.

The product of cumulative networks is cumulative itself.
For binary cumulative two-mode networks $\mathbf{A}$ and $\mathbf{B}$ the value
$c_{ij}(t)$ of the product $\mathbf{C} = \mathbf{A} \cdot \mathbf{B}$ is equal to the number
of different members of $P$ with which both $i$ and $j$ had contact in instants up to including the instant $t$.

\subsubsection{Temporal co-occurrence networks \label{bibtemp}}

Using the multiplication of temporal affiliation networks over the combinatorial semiring we
get the corresponding instantaneous and cumulative co-occurrence networks
\[  \mathbf{Ci} = \mathbf{Ai}^T \cdot \mathbf{Ai} \qquad \mbox{and}
    \qquad \mathbf{Cc} = \mathbf{Ac}^T \cdot \mathbf{Ac} \]      
           
The triple $(s,f,v)$ in a temporal quantity $ci_{pq}$ tells that in the time interval
$[s,f)$ there were $v$ events in which both $p$ and $q$ took part. 

The triple $(s,f,v)$ in a temporal quantity $cc_{pq}$ tells that in the time interval
$[s,f)$ there were in total $v$ accumulated events in which both $p$ and $q$ took part.

The diagonal (loop) weights  $ci_{pp}$ and  $cc_{pp}$ contain the temporal quantities
counting the number of events in the time intervals in which the participant
$p$ took part.

A typical example of such a network is the works authorship network $\mathbf{WA}$ where 
$E$ is the set of papers $W$, $P$ is the set of authors $A$, and $d$ is the publication
year.

\subsubsection{Example: Temporal coauthorship network  \label{excoau}} 

\begin{figure}
\begin{center}
\includegraphics[width=100mm]{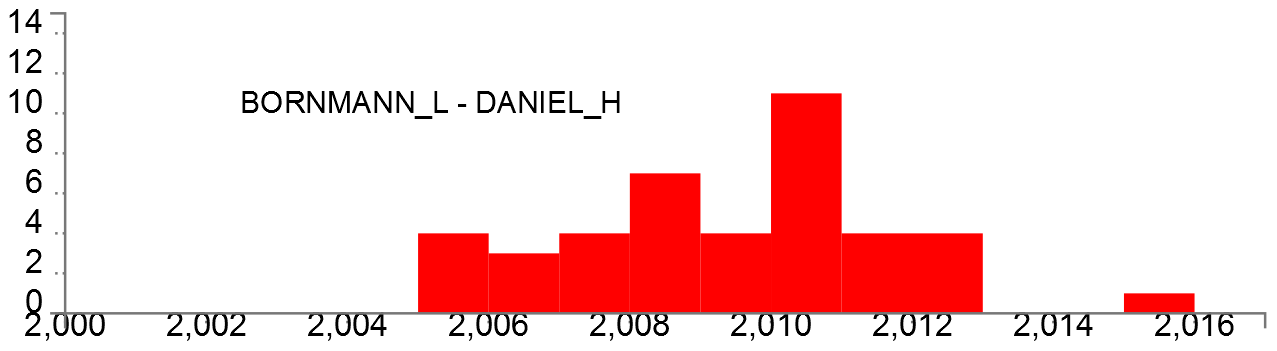}\\[2mm]
\includegraphics[width=100mm]{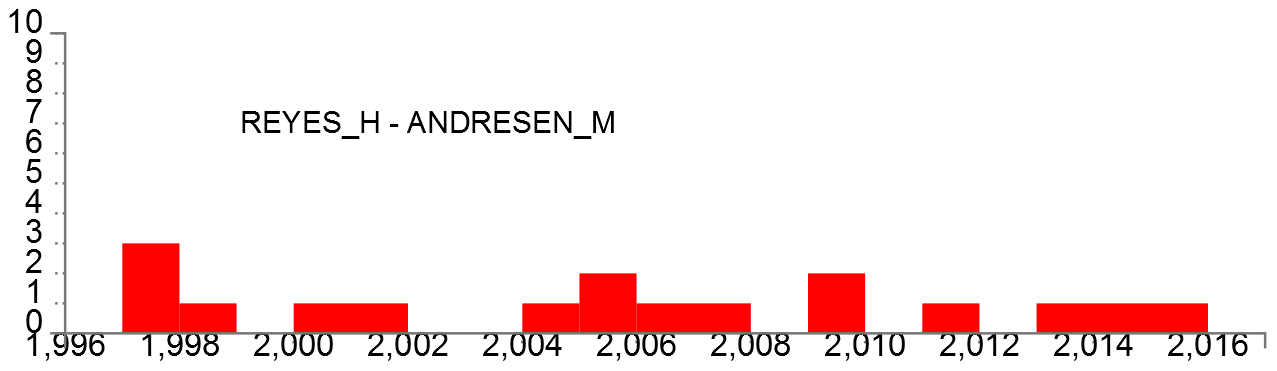}
\caption{Coauthorship\label{BR}}
\end{center}
\end{figure}

The  instantaneous coauthorship network $ \mathbf{Coi}$ is obtained as
\[ \mathbf{Coi} = \mathbf{WAi}^T \cdot \mathbf{WAi} \] 
Bibliographic networks are usually sparse. Often also the product of sparse networks is sparse itself. Considering in computation only non zero elements it can be computed fast \citep{OnBibl}. In our example, the network $\WA$ has 22104 works, 62106 authors and 80021 arcs. The derived network $\mathbf{Coi}$ has $633977$ edges and was computed on a laptop in 12.7 seconds.

For the peer review data we get the largest values 

$co[\mbox{BORNMANN\_L}, \mbox{DANIEL\_H}] = 42$,

$ co[\mbox{MOHER\_D}, \mbox{ALTMAN\_D}] = 24$,

$co[\mbox{REYES\_H}, \mbox{ANDRESEN\_M}] = 17$.

\noindent 
The corresponding temporal quantities $bd = tq(\mbox{BORNMANN\_L}, \mbox{DANIEL\_H})$ and $ra = tq(\mbox{REYES\_H},$ $ \mbox{ANDRESEN\_M})$ are
\begin{eqnarray*}
bd  & = & [(2005, 2006, 4), (2006, 2007, 3), (2007, 2008, 4), (2008, 2009, 7), (2009, 2010, 4),\\
     & &  (2010, 2011, 11), (2011, 2013, 4), (2015, 2016, 1)] \\
ra & = & [(1997, 1998, 3), (1998, 1999, 1), (2000, 2002, 1), (2004, 2005, 1), (2005, 2006, 2), \\
    & &  (2006, 2008, 1), (2009, 2010, 2), (2011, 2012, 1), (2013, 2016, 1)]
\end{eqnarray*}
Both temporal quantities are presented in Figure~\ref{BR}. The Python code is given in Appendix~A.4.

\subsubsection{Example: Temporal citations between journals  \label{excijo}} 

\begin{figure}
\begin{center}
\includegraphics[width=100mm]{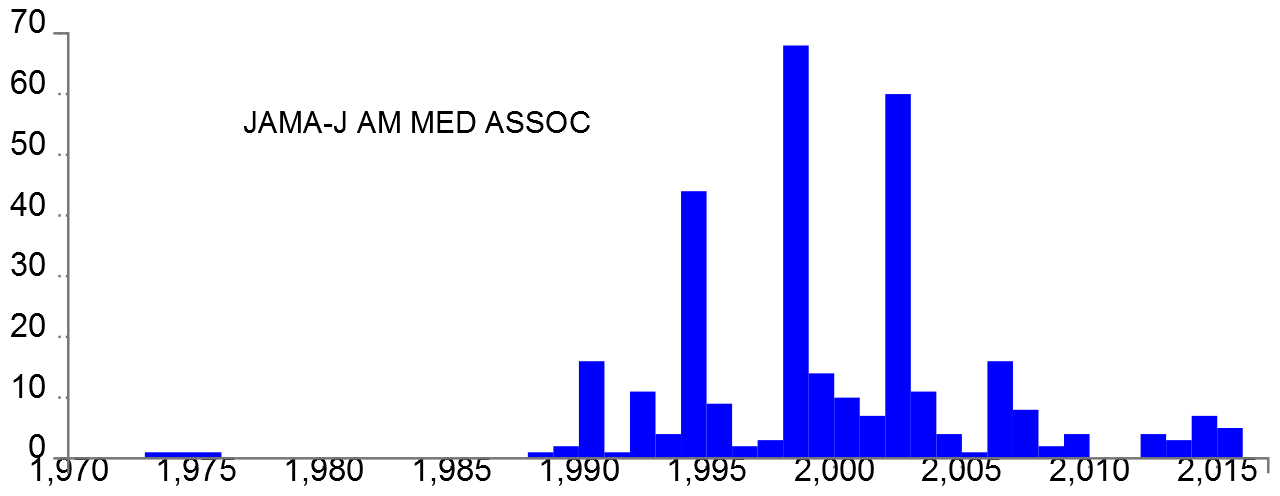}\\[2mm]
\includegraphics[width=100mm]{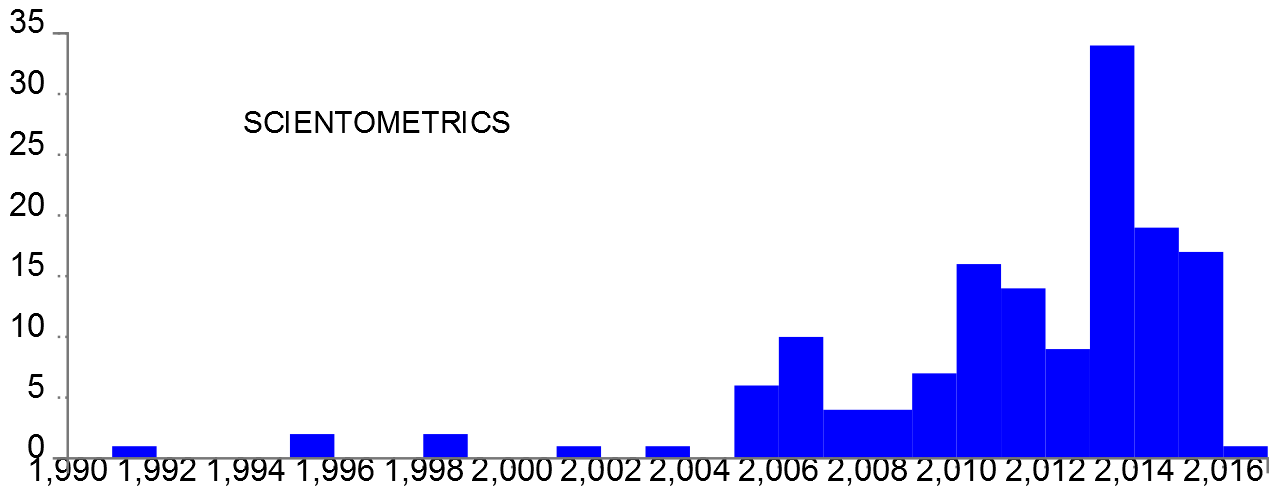}
\caption{Selfcitations\label{JS}}
\end{center}
\end{figure}

The derived network describing citations between journals is obtained as
\[ \mathbf{JCJ} = \WJ\mathbf{i}^T \cdot \Ci\mathbf{I} \cdot \WJ\mathbf{c} \]
Note that the third network in the product is cumulative.

The  weight of the element $jcj_{ij}$ is equal to the number of citations per year from works published in journal $i$ to
works published in journal $j$. In a special case when $i=j$ we get a temporal quantity describing selfcitations of journal $i$.
In the  peer review data the largest number of selfcitations are   320 in JAMA and 148 in Scientometrics. The corresponding temporal quantities $jm = jcj[\mbox{JAMA}, \mbox{JAMA}]$ and $sm = jcj[\mbox{SCIENTOMETRICS}, \mbox{SCIENTOMETRICS}]$ are:
\begin{eqnarray*}
jm  & = & [(1973, 1976, 1), (1988, 1989, 1), (1989, 1990, 2), (1990, 1991, 16), (1991, 1992, 1),\\
    & &       (1992, 1993, 11), (1993, 1994, 4), (1994, 1995, 44), (1995, 1996, 9), (1996, 1997, 2),\\
    & &       (1997, 1998, 3), (1998, 1999, 68), (1999, 2000, 14), (2000, 2001, 10), (2001, 2002, 7),\\
    & &       (2002, 2003, 60), (2003, 2004, 11), (2004, 2005, 4), (2005, 2006, 1), (2006, 2007, 16),\\
    & &      (2007, 2008, 8), (2008, 2009, 2), (2009, 2010, 4), (2012, 2013, 4), (2013, 2014, 3),\\
    & &      (2014, 2015, 7), (2015, 2016, 5)] \\
sm & = & [(1991, 1992, 1), (1995, 1996, 2), (1998, 1999, 2), (2001, 2002, 1), (2003, 2004, 1),\\
    & &       (2005, 2006, 6), (2006, 2007, 10), (2007, 2009, 4), (2009, 2010, 7), (2010, 2011, 16),\\
    & &       (2011, 2012, 14), (2012, 2013, 9), (2013, 2014, 34), (2014, 2015, 19), (2015, 2016, 17),\\
    & &      (2016, 2017, 1)]
\end{eqnarray*}
and are presented in Figure~\ref{JS}.

The largest number of citations are from journals  BMJ Open (142) and Scientometrics (108) to the unknown journal *****, followed by $bj = jcj[\mbox{BRIT MED J}, \mbox{JAMA-J AM MED ASSOC} ] $ and $pj = jcj[\mbox{PLOS ONE},$ $ \mbox{JAMA-J AM MED ASSOC} ] $ with totals 96 and 91.
\begin{figure}
\begin{center}
\includegraphics[width=100mm]{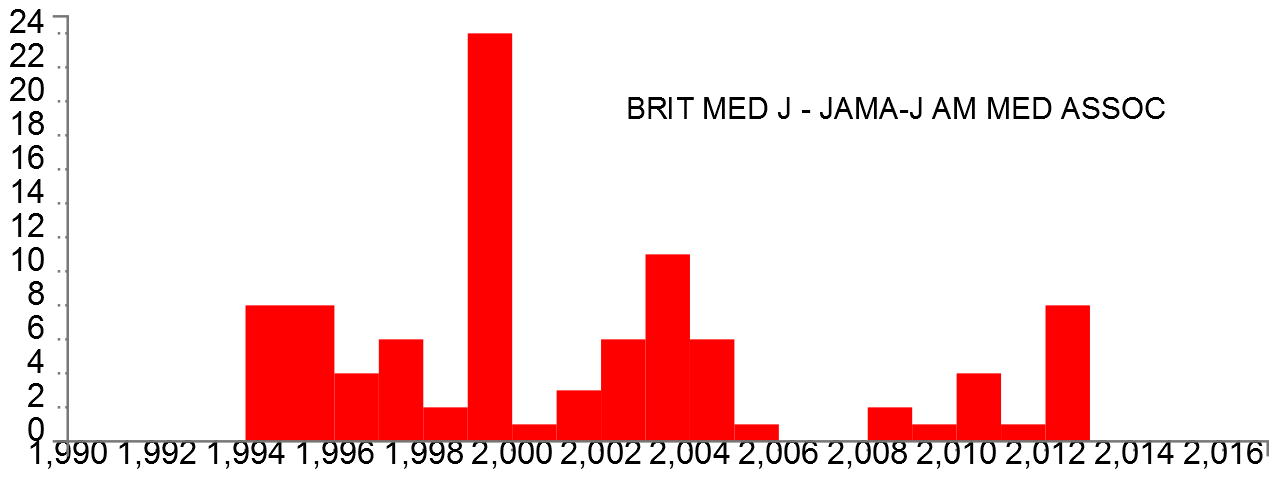}\\[2mm]
\includegraphics[width=100mm]{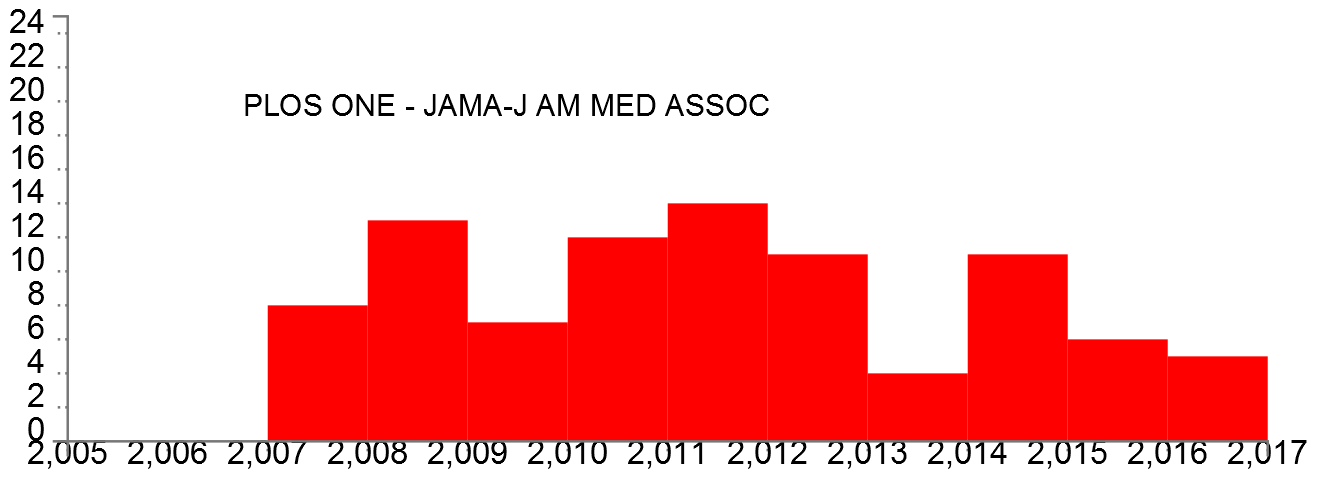}
\includegraphics[width=100mm]{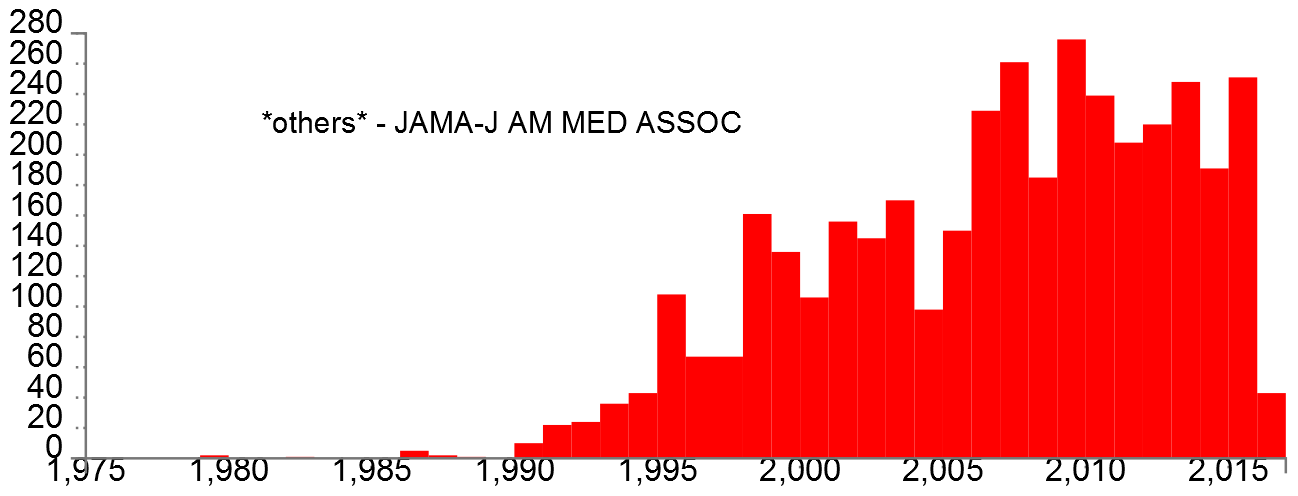}
\caption{Citations between journals\label{Jour}}
\end{center}
\end{figure}
\begin{eqnarray*}
bj  & = & [(1994, 1996, 8), (1996, 1997, 4), (1997, 1998, 6), (1998, 1999, 2), (1999, 2000, 24),\\
    & &       (2000, 2001, 1), (2001, 2002, 3), (2002, 2003, 6), (2003, 2004, 11), (2004, 2005, 6),\\
    & &       (2005, 2006, 1), (2008, 2009, 2), (2009, 2010, 1), (2010, 2011, 4), (2011, 2012, 1),\\
    & &       (2012, 2013, 8)] \\
pj & = & [(2007, 2008, 8), (2008, 2009, 13), (2009, 2010, 7), (2010, 2011, 12), (2011, 2012, 14),\\
    & &       (2012, 2013, 11), (2013, 2014, 4), (2014, 2015, 11), (2015, 2016, 6), (2016, 2017, 5)] 
\end{eqnarray*}
See the top and mid part of Figure~\ref{Jour}.

In the peer review data the journal JAMA is the most prominent. To get the temporal quantity describing citations of others to JAMA we compute $jci =  iS(\mathbf{JCJ},\mbox{JAMA-J AM MED ASSOC})$:
\begin{eqnarray*}    
 jci & = & [(1979, 1980, 2), (1982, 1983, 1), (1986, 1987, 5), (1987, 1988, 2), \\
    & &     (1988, 1989, 1),  (1990, 1991, 10), (1991, 1992, 22), (1992, 1993, 24),\\
    & &      (1993, 1994, 36), (1994, 1995, 43), (1995, 1996, 108), (1996, 1998, 67),\\
    & &       (1998, 1999, 161), (1999, 2000, 136), (2000, 2001, 106), (2001, 2002, 156), \\
    & &      (2002, 2003, 145), (2003, 2004, 170), (2004, 2005, 98), (2005, 2006, 150),\\
    & &       (2006, 2007, 229), (2007, 2008, 261), (2008, 2009, 185), (2009, 2010, 276),\\
    & &       (2010, 2011, 239), (2011, 2012, 208), (2012, 2013, 220), (2013, 2014, 248),\\ 
    & &     (2014, 2015, 191), (2015, 2016, 251), (2016, 2017, 43)]
\end{eqnarray*}
It is presented at the bottom of  Figure~\ref{Jour}. The Python code is given in Appendix~A.5.


Similarly we get the temporal network describing citations between authors
\[ \mathbf{ACA} = \WA\mathbf{i}^T \cdot \Ci\mathbf{I} \cdot \WA\mathbf{c} \]
The  weight of the element $aca_{ab}$ is equal to the number of citations per year from works coauthored by author $a$ to
works coauthored by author $b$.

\section{Conclusions}

We presented two ways (instantaneous and cumulative) to transform bibliographic networks, using the works' publication year, into corresponding temporal networks based on temporal quantities. They are a basis for a longitudinal approach to the analysis of temporal network which is an alternative to the traditional cross-sectional approach. Introducing a time dimension  can give additional insights into bibliographic networks. We also presented some methods for analyzing the obtained temporal networks and illustrated them with examples from  analysis of the peer review bibliography.

We presented only some examples to show that it works. The proposed approach can be extended in some directions:
\begin{itemize}
\item other node and link properties;
\item other derived networks combined with fractional approach;
\item normalization (proportions) of temporal properties considering the changes of the ``size'' of network through time;
\item clustering of temporal quantities to determine their types;
\item temporal networks methods produce large results. Special methods for identifying and presenting (visualizing) interesting
parts need to be developed.
\end{itemize}

\section*{Acknowledgments}
The paper is based on presentations on  1244. Sredin seminar, IMFM, Ljubljana, April 8, 2015; PEERE meeting, Vilnius,  March 7-9, 2017; and XXXVII Sunbelt workshop Beijing, China, May 30 -- June 4, 2017.
 
This work is supported in part by the Slovenian Research Agency (research program P1-0294 and research projects J1-9187, J7-8279 and BI-US/17-18-045), project PEERE (COST Action TD1306) and by Russian Academic Excellence Project '5-100'.

\newpage
\appendix
\section{Code in Nets}

\subsection{Converting Pajek net and clu files into temporal network in netsJSON}

To set up an environment for computing our examples we have to put in the directory \texttt{gdir} Python files (\texttt{Nets.py}, \texttt{TQ.py}, \texttt{search.py}, \texttt{coloring.py}, \texttt{IndexMinPQ.py}) from the library Nets, and in the subdirectory \texttt{cdir} the files  \texttt{TQchart.html}, \texttt{d3.v3.min.js} and \texttt{barData.js}.  The directory \texttt{ndir} contains the network data and  the directory \texttt{wdir} contains the results.
{
\footnotesize
\renewcommand{\baselinestretch}{0.8}
\begin{verbatim}
gdir = 'c:/path/Nets'
wdir = 'c:/path/Test/peere'
ndir = 'c:/path/WoS/peere2'
cdir = 'c:/path/Nets/chart'
import sys, os, datetime, json
sys.path = [gdir]+sys.path; os.chdir(wdir)
from TQ import *
from Nets import Network as N
net = ndir+"/WAd.net"
clu = ndir+"/Yeard.clu"
t1 = datetime.datetime.now(); print("started: ",t1.ctime(),"\n")
WAc = N.twoMode2netsJSON(clu,net,'WAcum.json',instant=False)
t2 = datetime.datetime.now() 
print("\nconverted to cumulative TN: ",t2.ctime(),"\ntime used: ", t2-t1)
WAi = N.twoMode2netsJSON(clu,net,'WAins.json',instant=True)
t3 = datetime.datetime.now() 
print("\nconverted to instantaneous TN: ",t3.ctime(),"\ntime used: ", t3-t2)
cit = ndir+"/CiteD.net"
Citei = N.oneMode2netsJSON(clu,cit,'CiteIns.json',instant=True)
t4 = datetime.datetime.now()
print("\nconverted to instantaneous TN: ",t4.ctime(),"\ntime used: ", t4-t3)
ia = WAi.Index()
ic = Citei.Index()
\end{verbatim}
}
\normalsize

\subsection{Productivities of authors}

{
\footnotesize
\renewcommand{\baselinestretch}{0.8}
\begin{verbatim}
>>> tit = 'BORNMANN_L';  b = ia[tit]
>>> pr = WAi.TQnetInSum(b)
>>> pr
[(2005, 2006, 4), (2006, 2007, 3), (2007, 2008, 4), (2008, 2009, 9),...
>>> TQ.TqSummary(pr)
(1900, 2017, 0, 14)
>>> TQmax = 15; Tmin = 1995; Tmax = 2016; w = 600; h = 150
>>> N.TQshow(pr,cdir,TQmax,Tmin,Tmax,w,h,tit,fill='red')
>>> cpr = WAc.TQnetInSum(b)
>>> cpr
[(2005, 2006, 4), (2006, 2007, 7), (2007, 2008, 11), (2008, 2009, 20),...
>>> TQmax = 65; Tmin = 1995; Tmax = 2016; w = 600; h = 250
>>> N.TQshow(cpr,cdir,TQmax,Tmin,Tmax,w,h,tit,fill='red')
>>> WAni = WAi.TQnormal()
>>> fpr = WAni.TQnetInSum(b)
>>> fpr
[(2006, 2007, 1.3333333333333333), (2007, 2008, 1.6666666666666665),...
>>> TQmax = 7; Tmin = 1995; Tmax = 2016; w = 600; h = 150
>>> N.TQshow(fpr,cdir,TQmax,Tmin,Tmax,w,h,tit,fill='red')
\end{verbatim}
}
\normalsize

\subsection{Citations between works}

{
\footnotesize
\renewcommand{\baselinestretch}{0.8}
\begin{verbatim}
>>> tit = 'PETERS_D(1982)5:187'; c = ic[tit]
>>> ci = Citei.TQnetInSum(c)
>>> ci
[(1982, 1983, 1), (1983, 1984, 4), (1984, 1986, 3), (1986, 1987, 2), ...
>>> TQmax = 15; Tmin = 1980; Tmax = 2016; w = 600; h = 150
>>> N.TQshow(ci,cdir,TQmax,Tmin,Tmax,w,h,tit,fill='blue')
>>> tit = 'HIRSCH_J(2005)102:16569'; c = ic[tit]
>>> ci = Citei.TQnetInSum(c)
>>> ci
[(2005, 2006, 0), (2006, 2007, 3), (2007, 2008, 4), (2008, 2009, 7), ...
>>> TQmax = 25; Tmin = 2000; Tmax = 2017; w = 600; h = 250
>>> N.TQshow(ci,cdir,TQmax,Tmin,Tmax,w,h,tit,fill='blue')
\end{verbatim}
}
\normalsize

\subsection{The most important  journals\label{important}}


{
\footnotesize
\renewcommand{\baselinestretch}{0.8}
\begin{verbatim}
>>> jrn = ndir+"/WJd.net"
>>> WJc = N.twoMode2netJSON(clu,jrn,'WJcum.json',instant=False)
>>> WJi = N.twoMode2netJSON(clu,jrn,'WJins.json',instant=True)
>>> J = list(WJi.nodesMode(2))
>>> Jt = [ (j, WJi._nodes[j][3]['lab'], TQ.cutGT(WJi.TQnetInSum(j),0)) for j in J ]
>>> p = [0,1971,1981,1991,2001,2006,2011,2016,3000]
>>> Jr = [ (j,l,TQ.changeTime(a,p)) for (j,l,a) in Jt ]
>>> I = { Jt[j][1] : j for j in range(len(Jt)) }
>>> JL = [ "BEHAV BRAIN SCI", "BMJ OPEN", "BRIT MED J", "CUTIS",
   "J ASSOC OFF AGR CHEM", "JAMA-J AM MED ASSOC", "J SEX MED",
   "LANCET", "MED J AUSTRALIA", "NATURE", "NEW ENGL J MED",
   "PLOS ONE", "SCIENCE", "SCIENTOMETRICS" ]
>>> IJ = [ I[j] for j in JL ]; Ir = [ Jr[i] for i in IJ]
\end{verbatim}
}
\normalsize\noindent
In the library TQ we included a new function \texttt{changeTime} that recodes a temporal quantity $a$ into new time intervals determined by a sequence $p$.

\subsection{Temporal coauthorship network}

{
\footnotesize
\renewcommand{\baselinestretch}{0.8}
\begin{verbatim}
>>> Co = WAi.TQtwo2oneCols()
>>> Co.saveNetsJSON('CoIns.json',indent=2)
>>> Co.delLoops()
>>> C = Co.TQtopLinks(thresh=15)
>>> tit = C[0][2]+' - '+C[0][3]; bd = C[0][5]
>>> TQmax = 15; Tmin = 2000; Tmax = 2017; w = 600; h = 150
>>> N.TQshow(bd,cdir,TQmax,Tmin,Tmax,w,h,tit,fill='red')
>>> tit = C[2][2]+' - '+C[2][3]; ra = C[2][5]
>>> TQmax = 10; Tmin = 1996; Tmax = 2017; w = 600; h = 150
>>> N.TQshow(ra,cdir,TQmax,Tmin,Tmax,w,h,tit,fill='red')
>>> TQ.total(bd), TQ.total(ra)
(42, 17)
\end{verbatim}
}
\normalsize

\subsection{Citations between journals}

{
\footnotesize
\renewcommand{\baselinestretch}{0.8}
\begin{verbatim}
>>> JCJ = N.TQmultiply(N.TQmultiply(WJi.transpose(),Citei.one2twoMode()),WJc,True)
>>> L = JCJ.TQtopLoops(thresh=100)
>>> tit = L[0][1]; jm = L[0][3]
>>> TQmax = 70; Tmin = 1970; Tmax = 2017; w = 600; h = 200
>>> N.TQshow(jm,cdir,TQmax,Tmin,Tmax,w,h,tit,fill='blue')
>>> tit = L[1][1]; sm = L[1][3]
>>> TQmax = 35; Tmin = 1990; Tmax = 2017; w = 600; h = 200
>>> N.TQshow(sm,cdir,TQmax,Tmin,Tmax,w,h,tit,fill='blue')
>>> JCJ.delLoops()
>>> T = JCJ.TQtopLinks(thresh=70)
>>> tit = T[2][2]+' - '+T[2][3]; bj = T[2][5]
>>> TQmax = 25; Tmin = 1990; Tmax = 2017; w = 600; h = 200
>>> N.TQshow(bj,cdir,TQmax,Tmin,Tmax,w,h,tit,fill='red')
>>> tit = T[3][2]+' - '+T[3][3]; pj = T[3][5]
>>> TQmax = 25; Tmin = 2005; Tmax = 2017; w = 600; h = 200
>>> N.TQshow(pj,cdir,TQmax,Tmin,Tmax,w,h,tit,fill='red')
>>> jci = TQ.cutGE(JCJ.TQnetInSum(T[2][1]),1e-10)
>>> TQ.TqSummary(jci)
(1979, 2017, 1, 276)
>>> TQ.total(jci)
3861
>>> tit = '*others* - '+T[2][3]
>>> TQmax = 280; Tmin = 1975; Tmax = 2017; w = 600; h = 200
>>> N.TQshow(jci,cdir,TQmax,Tmin,Tmax,w,h,tit,fill='red')
\end{verbatim}
}
\normalsize

\end{document}